\documentstyle[epsf,twocolumn,prl,aps]{revtex}
\begin{document}
%\draft

\title{Disorder-Induced Critical Phenomena in Hysteresis:\\
Numerical Scaling in Three and Higher Dimensions}

\author{Olga Perkovi\'{c}$^*$, Karin A. Dahmen$^\dagger$, and James P. Sethna}

\address{Laboratory of Atomic and Solid State Physics, Cornell
  University, Ithaca, NY 14853-2501\hfil\break
  $^*$ Current address: McKinsey \& Company, 555 California Street, Suite 4700,
  San Francisco, California 94104\hfil\break
  $^\dagger$ Current address: Department of Physics, Harvard University,
  Cambridge, MA 02138
}

\date{\today}

\maketitle

\begin{abstract}
We present numerical simulations of avalanches and critical phenomena associated
with hysteresis loops, modeled using the zero--temperature random--field Ising
model. We study the transition between smooth hysteresis loops and loops with
a sharp jump in the magnetization, as the disorder in our model is decreased.
In a large region near the critical point, we find scaling and critical phenomena,
which are well described by the results of an $\epsilon$ expansion about six
dimensions. We present the results of simulations in 3, 4, and 5 dimensions,
with systems with up to a billion spins ($1000^3$).

\end{abstract}

\pacs{PACS numbers: 05.70.Jk,75.10.Nr,75.40.Mg,75.60.Ej}

\section{Introduction}

The increased interest in real materials in condensed matter physics has
brought disordered systems into the spotlight. Dirt changes the free
energy landscape of a system, and can introduce metastable states with
large energy barriers~\cite{barrier}. This can lead to extremely slow
relaxation towards the equilibrium state. On long length scales and
practical time scales, a system driven by an external field will move
from one metastable local free-energy minimum to the next. The
equilibrium, global free energy minimum and the thermal fluctuations
that drive the system toward it, are in this case irrelevant. The state
of the system will instead depend on its history.

The motion from one local minima to the next is a collective process
involving many local (magnetic) domains in a local region - {\it an
avalanche}. In magnetic materials, as the external magnetic field $H$ is
changed continuously, these avalanches lead to the magnetic noise: the
Barkhausen effect\cite{Jiles,McClure}. This effect can be picked up as
voltage pulses in a coil surrounding the magnet. The distribution of
pulse (avalanche) sizes is
found~\cite{McClure,Barkhausen,Urbach,SOC_example} to follow a power law
with a cutoff after a few decades, and was interpreted by
some~\cite{SOC_example} to be an example of self-organized
criticality (SOC)~\cite{SOC}. (In SOC, a system organizes itself into a
critical state without the need to tune an external parameter.) Other
systems can exhibit avalanches as well. Several examples where disorder
may play a part are: superconducting vortex line
avalanches~\cite{Field}, resistance avalanches in superconducting
films~\cite{Wu}, and capillary condensation of helium in
Nuclepore~\cite{Nuclepore}.

The history dependence of the state of the system leads to hysteresis.
Experiments with magnetic tapes~\cite{Berger} have shown that the shape
of the hysteresis curve changes with the annealing temperature. The
hysteresis curve goes from smooth to discontinuous as the annealing
temperature is increased. This transition can be explained in terms of a
{\it plain old critical point} with two tunable parameters: the
annealing temperature and the external field. At the critical
temperature and field, the correlation length diverges, and the
distribution of pulse (avalanche) sizes follows a power law.

We have argued earlier~\cite{prl3} that the Barkhausen noise
experiments can be quantitatively explained by a model~\cite{prl1}
with two tunable parameters (external field and disorder), which
exhibits {\it universal}, non-equilibrium collective behavior. The model
is athermal and incorporates collective behavior through nearest
neighbor interactions. The role of {\it dirt} or disorder, as we call
it, is played by random fields. This paper presents the results and
conclusions of a large scale simulation of that model: the
non-equilibrium zero-temperature Random Field Ising Model (RFIM), with a
deterministic dynamics. The results compare very well to our $\epsilon$
expansion\cite{Dahmen1,Dahmen2}, and to experiments in Barkhausen
noise\cite{prl3}. 

We should mention that there are other models for avalanches in
disordered magnets. There is a large body of work on depinning
transitions and the motion of the single
interface\cite{Robbins,depinning,SameDepinning}. In these models,
avalanches occur only at the growing interface. Our model though, deals
with many interacting interfaces: avalanches can grow anywhere in the
system. Models of hysteresis similar to ours exist~\cite{Bertram}, including
ones with random bonds\cite{RandomBonds,Vives} and random anisotropies.

This paper is a condensed version of an unpublished manuscript,
available electronically~\cite{OlgaLong}. We focus here on the numerical
results and scaling methods in dimensions three through six. Some of the
other topics touched upon in the original manuscript are being published
separately. Our interpretation of the behavior in dimension two has been
substantially altered by further analysis~\cite{Kuntz}. A full
description of the numerical method is available, including sample code
and executables on the Web~\cite{Numerical}. For a full discussion of
the behavior in mean field theory, and interesting behavior below the
critical point in seven and nine dimensions, we refer the reader to the
electronic version of the original manuscript~\cite{OlgaLong}, and to
recent work on the Bethe lattice~\cite{Dhar}.

\section{The Model}

The model we use is the zero--temperature random--field Ising
model~\cite{Bertram,Robbins,prl1,prl3}, which we briefly review here.
Magnetic domains are represented by spins $s_i$ on
a hypercubic lattice, which can take two values: $s_i = \pm 1$. The
spins interact ferromagnetically with their nearest neighbors with a
strength $J_{ij}$, and are exposed to a uniform magnetic field $H$
(which is directed along the spins). Disorder is simulated by a random field
$h_i$, associated with each site of the lattice, which is given by a
gaussian distribution function $\rho (h_i)$:
\begin{equation}
\rho (h_i) = {1 \over {\sqrt {2\pi}} R}\ e^{-{h_i}^2 \over 2R^2}
\label{eq:model_1}
\end{equation}
of width proportional to $R$, which we call the disorder. The Hamiltonian is then
\begin{equation}
{\cal H} = - \sum_{<i,j>} J_{ij} s_i s_j - \sum_{i} (H + h_i) s_i
\label{eq:model_2}
\end{equation}
For the analytic calculation, as well as the simulation, we have set the
interaction between the spins to be independent of the spins and equal
to one for nearest neighbors, $J_{ij}=J=1$, and zero otherwise. We use
periodic boundary conditions in the results of this paper; we've checked
that the results near $R_c$ are unchanged when a slab of pre--flipped spins 
is introduced (fixed boundary conditions along two sides).

The dynamics is deterministic, and is defined such that a spin $s_i$
will flip only when its local effective field $h^{ef\!f}_i$:
\begin{equation}
h^{ef\!f}_i = J \sum_{j} s_j + H + h_i
\label{eq:model_3}
\end{equation}
changes sign. All the spins start pointing down ($s_i=-1$ for all $i$).
As the field is adiabatically increased, a spin will flip. Due to the
nearest neighbor interaction, a flipped spin will push a neighbor to
flip, which in turn might push another neighbor, and so on, thereby
generating an avalanche of spin flips. During each avalanche, the
external field is kept constant. For large disorders, the distribution
of random fields is wide, and spins will tend to flip independently of
each other. Only small avalanches will exist, and the magnetization
curve will be smooth. On the other hand, a small disorder implies a
narrow random field distribution which allows larger avalanches to
occur. As the disorder is lowered, at the disorder $R=R_c$ and field
$H=H_c$, an infinite avalanche in the thermodynamic system will occur
for the first time, and the magnetization curve will show a
discontinuity. Near $R_c$ and $H_c$, we find critical scaling behavior
and avalanches of all sizes. Therefore, the system has two tunable
parameters: the external field $H$ and the disorder $R$. We found from
the mean field calculation~\cite{Dahmen1,Dahmen2} and the simulation
that a discontinuity in the magnetization exists for disorders $R \le
R_c$, at the field $H_c(R) \ge H_c(R_c)$, but that only at $(R_c, H_c)$,
do we have critical behavior. For finite size systems of length $L$, the
transition occurs at the disorder $R_c^{ef\!f}(L)$ near which avalanches
first begin to span the system in one of the {\it d} dimensions
(spanning avalanches). The effective critical disorder $R_c^{ef\!f}(L)$
is larger than $R_c$, and $R_c^{ef\!f}(L) \rightarrow R_c$ as $L
\rightarrow \infty$.

The algorithm we use to simulate this model is described in a separate
manuscript~\cite{Numerical}. For a simulation with $N$ spins, the computer
time scales as $N \log N$ and the memory required for the simulation
scales to one bit per spin ({\it i.e.}, we do not store the random fields). 

\section{Scaling}

We use data obtained from the simulation to find and describe the
critical transition. We do so using {\bf scaling collapses}, which we review
briefly here.
For example, the magnetization as a function of external field $H$
is expected to have the form
\begin{equation}
M(H,R) - M_c(H_c,R_c) \sim |r|^\beta\ {\cal M}_{\pm}(h'/|r|^{\beta\delta})
\label{eq:scaling_1}
\end{equation}
where $M_c$ is the critical magnetization (the magnetization at $H_c$,
for $R=R_c$), $r=(R-R_c)/R$ and $h=(H-H_c)$ are the reduced disorder and
reduced field respectively, 
\begin{equation}
h' = h+B r
\label{eq:scaling_2}
\end{equation}
is a (non-universal) rotation between
the experimental control variables $(r,h)$ and the scaling variables $(r,h')$,
and ${\cal M}_{\pm}$ is a universal scaling
function ($\pm$ refers to the sign of $r$).\footnote{In the plots shown
in this paper, we use $r=(R-R_c)/R$, which we have found 
produces better collapses than using $r=(R-R_c)/R_c$. The latter
is more traditional, but the two definitions agree as $R\to R_c$, and differ
by an amount which is irrelevant in a renormalization--group sense. One method
we use to estimate error in our exponents is to compare extrapolations based
on the two definitions.}
Scaling is expected asymptotically
for small $r$ and $h$ --- {\it i.e.}, for $H$ near $H_c$ and $R$ near $R_c$. 
The critical exponent $\beta$ gives the scaling for the magnetization at
the critical field $H_c$ ($h=0$). If we plot
$|r|^{-\beta}  (M(H,R) - M_c(H_c,R_c)$ versus $h/r^{\beta\delta}$,
we should obtain the curve ${\cal M}(x)$, independently of what disorder
$R$ we choose (so long as it is close to $R_c$): different experimental and
numerical data sets should collapse onto one universal curve ${\cal M}(x)$.
(Actually, one has two curves ${\cal M}_\pm$ depending on whether $R>R_c$
or $R<R_c$.) We use scaling forms similar to (\ref{eq:scaling_1}) to
analyze all of our measurements.

One can easily show using the scaling form (\ref{eq:scaling_1}) that the
magnetization scales with a power law $M-M_c \sim h^\delta$ at $R_c$,
and that the jump in the magnetization (the size of the infinite avalanche)
scales as $\Delta M \sim r^\beta$ as one varies the disorder below $R_c$.
Thus the critical exponents $\beta$ and $\delta$ give the power laws for
the singularities in these measured quantities: indeed, that is how these
exponents were originally defined and measured. In our system, we will find
that directly measuring power laws is not effective in getting good exponents:
the critical regime is so large that we need both to use the general scaling
form and to extrapolate to the critical point.

The explanatory power of the theory resides in the fact that the same
universal critical exponents $\beta$ and $\delta$ and the same universal
function ${\cal M}(x)$ should be obtained by simulations at different
values of the disorder, simulations of different Hamiltonians, and
simulations of real experiments, so long as the systems share certain
important features and symmetries (so long as they lie in the same
universality class). The underlying explanation for why universality and
scaling should occur near the critical point is given by the
renormalization group~\cite{RG,Dahmen1,Dahmen2,OlgaLong}. Above six
dimensions, fluctuations are asymptotically not important, and we can
calculate ${\cal M}(x)$ and the values of $\beta$ and $\delta$ from mean
field theory ($\beta_{\rm MF} = 1/2$, $\delta_{\rm MF} = 3$~\cite{prl1}.
Below six dimensions, the exponents and scaling curves are non-trivial,
and to find them one must rely on either perturbative
methods~\cite{Dahmen1,Dahmen2}, experiments, or numerical
methods~\cite{prl1,prl3} as used here.

\section{The Simulation Results}

The following measurements were obtained from the simulation as a function
of disorder R: \par
$\bullet$ the magnetization $M(H,R)$ as a function of the \par external
field $H$. \par
$\bullet$ the avalanche size distribution integrated over the \par
field $H$: $D_{int}(S,R)$. \par
$\bullet$ the avalanche correlation function integrated over the \par
field $H$:
$G_{int}(x,R)$. \par
$\bullet$ the number of spanning avalanches $N(L,R)$ as a \par function of the
system length $L$, integrated over the \par field $H$. \par
$\bullet$ the discontinuity in the magnetization $\Delta M (L,R)$ as \par a
function of the system length $L$. \par
$\bullet$ the second $\langle S^2 \rangle_{int}(L,R)$, third
$\langle S^3 \rangle_{int}(L,R)$, and \par
fourth $\langle S^4 \rangle_{int}(L,R)$
moments of the avalanche size \par
distribution as a function of the system length $L$, \par
integrated over the field $H$. \par

{\noindent In addition, we have measured:} \par
$\bullet$ the avalanche size distribution $D(S,H,R)$ as a \par function of the
field $H$ and disorder $R$. \par
$\bullet$ the distribution of avalanche times $D_{t}^{(int)}(S,t)$
as a \par function of the avalanche size $S$, at
$R=R_c$, integrated \par over the field $H$. \par

\subsection{Magnetization Curves}

Unfortunately the most obvious measured quantity in our simulations,
the magnetization curve $M(H)$, is the one which collapses least well in our
simulations. We start with it nonetheless.

\begin{figure}[thb]
  \begin{center}
    \epsfxsize=8cm
    \epsffile{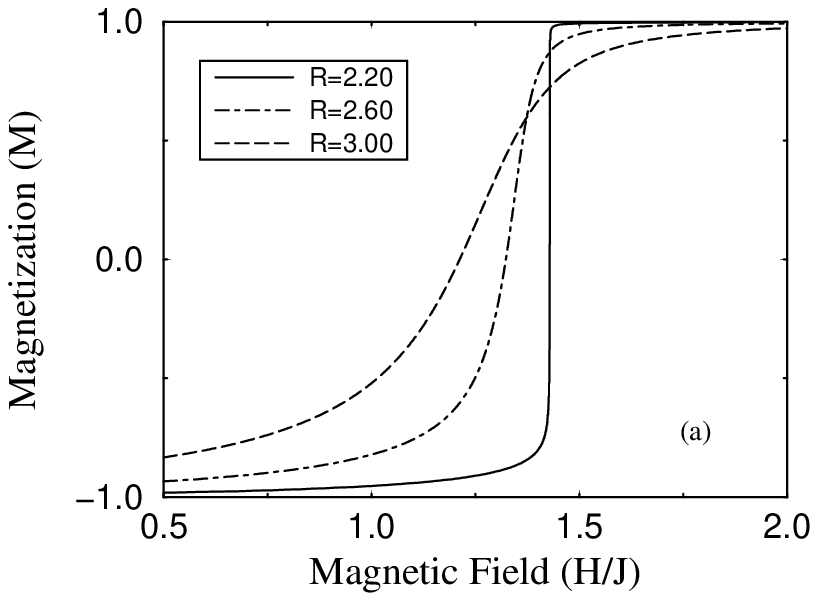}
    \epsfxsize=8cm
    \epsffile{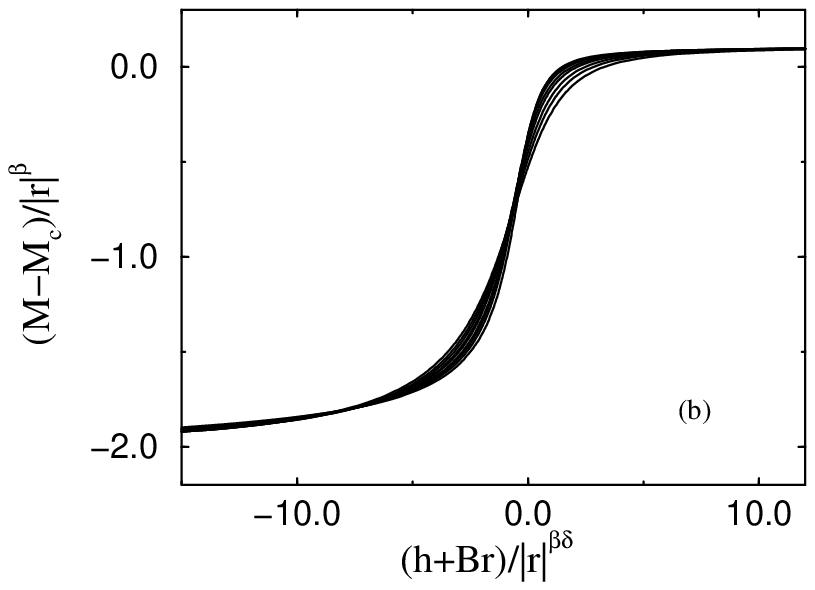}
  \end{center}
\caption{(a) {\bf Magnetization curves in $3$ dimensions} for size
$L=320$, and three values of disorder. The curves are averages of up to
$48$ different random field configurations. Note the discontinuity in
the magnetization for $R=2.20$. In finite size systems, the
discontinuity in the magnetization curve occurs even for $R>R_c$
($R_c=2.16$ in $3$ dimensions). (b) Scaling collapse (see
text) of the magnetization curves in $3$ dimensions for size $L=320$.
The disorders range from $R=2.35$ to $R=3.20$. The critical
magnetization is chosen as $M_c=0.9$ from an analysis of the
magnetization curves and is kept fixed during the collapse. The universal
exponents are $\beta=0.036$, $\beta\delta=1.81$. The non--universal
critical field $H_c = 1.435$, critical disorder $R_c = 2.16$, and 
rotation parameter $B=0.39$.
}
\label{fig:3d_MofH}
\end{figure}

Figure \ref{fig:3d_MofH}a shows the magnetization curves obtained from
our simulation in $3$ dimensions for several values of the disorder $R$.
As the disorder $R$ is decreased, a discontinuity or jump in the
magnetization curve appears where a single avalanche occupies a large
fraction of the total system. In the thermodynamic limit this would be
the infinite avalanche: the largest disorder at which it occurs is the
critical disorder $R_c$. For finite size systems, like the ones we use
in our simulation, we observe an avalanche which spans the system at a
higher disorder, which gradually approaches $R_c$ as the system size
grows.

\begin{figure}
  \begin{center}
    \epsfxsize=8cm
    \epsffile{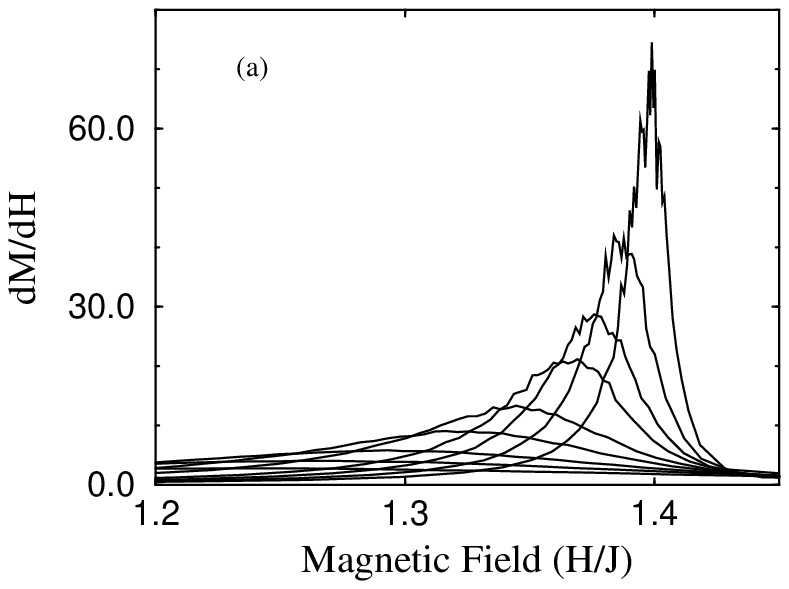}
    \epsfxsize=8cm
    \epsffile{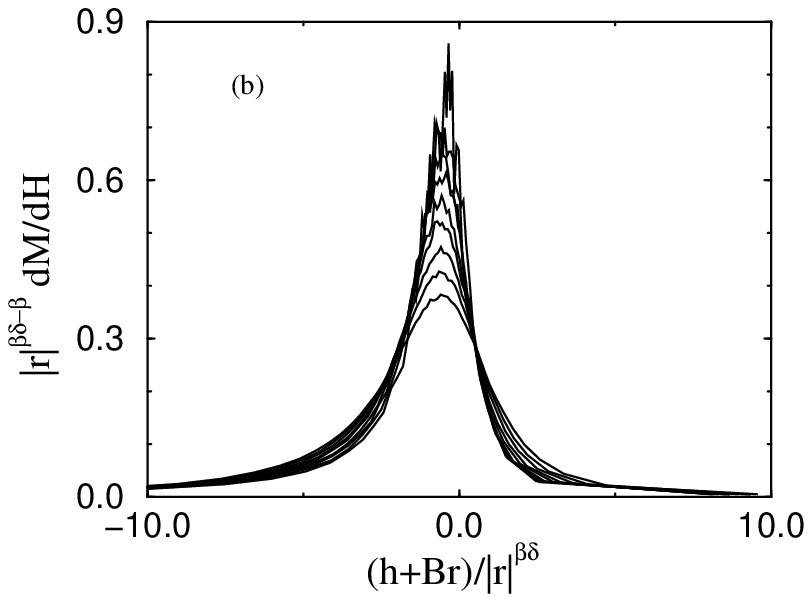}
  \end{center}
\caption{{\bf $dM/dH$ curves in $3$ dimensions}
(a)~Derivative of the magnetization $M$ with respect to the field $H$
for disorders $R=$ $2.35$, $2.4$, $2.45$, $2.5$, $2.6$, $2.7$,
$2.85$, $3.0$, and $3.2$ (highest to lowest peak), 
(b)~Scaling collapse of the data in (a) with $\beta =0.036$,
$\beta\delta = 1.81$, $B=0.39$, $H_c= 1.435$, and $R_c=2.16$. While the
curves are not collapsing onto a single curve, the quality of the
collapse is quite similar to that found at similar distances from $R_c$
in mean field theory~\protect\cite{OlgaLong}, for which we know analytically
that scaling works as $R \to R_c$.
}
\label{fig:dMdH_3d}
\end{figure}

Figure~\ref{fig:dMdH_3d} shows the slope $dM/dH$ and its scaling collapse.
By using this derivative, the critical region is emphasized as the peak
in the curve, and the dependence on the parameter $M_c$ drops out.
The lower graphs in figures~\ref{fig:3d_MofH}b and~\ref{fig:dMdH_3d}b
show the scaling collapses of the magnetization and its slope. Clearly
in neither case is all the data collapsing onto a single curve. This
would be distressing, were it not for the fact that this also occurs in
mean field theory~\cite{OlgaLong} at a similar distance to the critical point.

Because the scaling of the magnetization is so bad, we use other quantities
to estimate the critical exponents and the location of the critical point
(tables~\ref{table:exponents} and~\ref{table:RH}). Fixing these quantities,
we use the collapse of the $dM/dH$ curves to extract the rotation $B$ mixing
the experimental variables $r$ and $h$ into the scaling variable $h'=h + B r$
(equation~\ref{eq:scaling_2}).

\subsection{Avalanche Size Distribution}

\subsubsection{Integrated Avalanche Size Distribution}

\begin{figure}
  \begin{center}
    \epsfxsize=8cm
    \epsffile{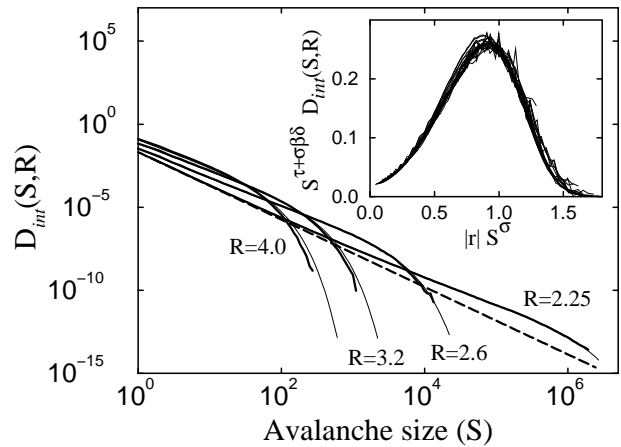}
  \end{center}
\caption{
{\bf Integrated avalanche size distribution curves in $3$ dimensions}
for $320^3$ spins and disorders $R=4.0, 3.2$, and $2.6$. The last curve
is at $R=2.25$, for a $1000^3$ spin system. The $320^3$ curves are
averages over up to $16$ initial random field configurations. The inset
shows the scaling collapse of the integrated avalanche size distribution
curves in $3$ dimensions, using $r=(R-R_c)/R$,
$\tau+\sigma\beta\delta=2.03$, and $\sigma=0.24$, for sizes $160^3$,
$320^3$, $800^3$, and $1000^3$, and disorders ranging from $R=2.25$ to
$R=3.2$ ($R_c = 2.16$). The two top curves in the collapse, at $R=3.2$,
show noticeable corrections to scaling. The thick dark curve through the
collapse is the fit to the data (see text). In the main figure, the
distribution curves obtained from the fit to the collapsed data are
plotted (thin lines) alongside the raw data (thick lines). The straight
dashed line is the expected asymptotic power law behavior: $S^{-2.03}$,
which does not agree with the measured slope of the raw data due to the
shape of the scaling function (see text).
}
\label{fig:aval_3}
\end{figure}

In our model the spins flip in avalanches: each spin can kick over one or
more neighbors in a cascade.  These
avalanches come in different sizes. The integrated avalanche size
distribution is the size distribution of all the avalanches that occur
in one branch of the hysteresis loop (for $H$ from $-\infty$ to
$\infty$). Figure~\ref{fig:aval_3}\cite{prl3} shows some of the raw
data (thick lines) in $3$ dimensions.   Note that the curves follow an
approximate power law behavior over several decades. Even $50\%$ away from
criticality (at $R=3.2$), there are still two decades of scaling, which
implies that the critical region is large. In experiments, a few decades
of scaling could be interpreted in terms of self-organized criticality
(SOC). However, our model and simulation suggest that several decades of
power law scaling can still be present rather ${\it far}$ from the
critical point (note that the size of the critical region is
non--universal). The slope of the log-log avalanche size distribution at
$R_c$ gives the critical exponent $\tau + \sigma \beta \delta$. Notice,
however, that the apparent slopes in figure~\ref{fig:aval_3} continue to
change even after several decades of apparent scaling is obtained.
The cutoff in the power law diverges as the critical disorder $R_c$ is
approached. This cutoff size scales as $S \sim |r|^{-1/\sigma}$.

These critical exponents can be obtained by using a scaling collapse for
the curves of figure~\ref{fig:aval_3}, shown in the inset. 
The scaling form is
\begin{equation}D_{int}(S,R) \sim S^{-(\tau+\sigma\beta\delta)}\
\bar{{\cal D}}^{(int)}_{+} (S^{\sigma}|r|)
\label{eq:aval_3d}
\end{equation}
where $\bar{{\cal D}}_{+}^{(int)}$ is the scaling function (the $+$ sign
indicates that the collapsed curves are for $R > R_c$). We are sufficiently
far from the critical point that corrections to scaling are important:
as described in reference~\cite{OlgaLong}, we do collapses for small ranges
of $R$ and then linearly extrapolate the best--fit critical exponents to $R_c$.
We estimate from this curve that the critical
exponents $\tau+\sigma\beta\delta=2.03$ and $\sigma=0.24$

The scaling function ${\cal D}_{+}^{(int)}(X)$ with $X=S^{\sigma}|r|$
is a universal prediction of our model. To facilitate comparisons with
experiments, we fit a curve to the data collapse in the inset of
figure~\ref{fig:aval_3}. We have fit the scaling collapses in dimensions
$3$, $4$, and $5$ to a phenomenological form of an exponential times
a polynomial. In three dimensions, our fit is 
\begin{eqnarray}
\bar{{\cal D}}_{+}^{\it (int)}(X)\ =\ e^{-0.789X^{1/\sigma}}\ \times
\nonumber \\
(0.021+0.002X+0.531X^2-0.266X^3+0.261X^4)
\label{eq:aval_fit_3d}
\end{eqnarray}
where $1/\sigma=4.20$.  The
distribution curves obtained using the above fit are plotted (thin lines
in figure~\ref{fig:aval_3}) alongside the raw data (thick lines). They
agree remarkably well even far above $R_c$. We should recall though,
that the fitted curve to the collapsed data can differ from the ``real''
scaling function even for large sizes and close to the critical disorder (in
mean field~\cite{OlgaLong} the error in the corresponding curve was about
$10\%$).

\begin{figure}
  \begin{center}
    \epsfxsize=8cm
    \epsffile{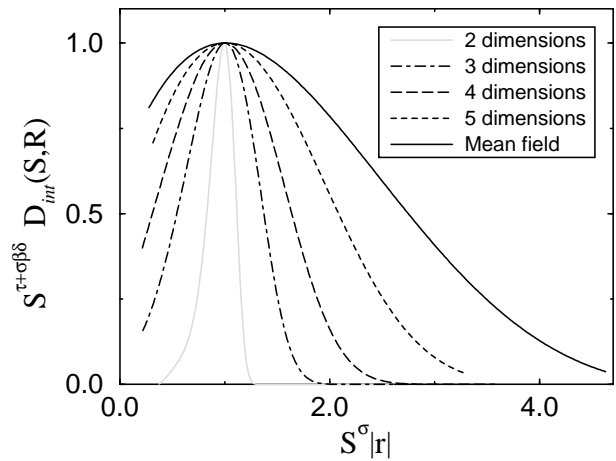}
  \end{center}
\caption{{\bf Integrated avalanche size distribution scaling functions
in $2$, $3$, $4$, and $5$ dimensions, and mean field.} The curves are
fits (see text) to the scaling collapses done with exponents from
Table~\protect\ref{table:exponents} and corresponding calculations in
two dimensions~\protect\cite{Kuntz}. The peaks are aligned to fall on (1,1).
Due to the ``bump'' in the scaling function the power law exponent can
not be extracted from a linear fit to the raw data for reasonable simulation
sizes~\protect\cite{prl3}.
}
\label{fig:bump_345}
\end{figure}

The scaling function in the inset of figure~\ref{fig:aval_3} has a
peculiar shape: it grows by a factor of ten before cutting off. The
consequence of this bump in the shape is that in the simulations it takes many
decades in the size distribution for the slope to converge to the
asymptotic power law. This can be seen from the comparison between a
straight line fit through the $R=2.25$ (billion spin) simulation in 
figure~\ref{fig:aval_3} and the asymptotic power law $S^{-2.03}$ obtained from
extrapolating the scaling collapses (thick dashed straight line in
the same figure). A similar bump exists in other dimensions and mean
field as well. Figure~\ref{fig:bump_345} shows the scaling functions in
different dimensions and in mean field. In this graph, the scaling
functions are normalized to one and the peaks are aligned (the scaling
forms allow this). The curves plotted in figure~\ref{fig:bump_345} are
not raw data but fits to the scaling collapse in each dimensions, as was
done in the inset of figure~\ref{fig:aval_3}. For $5$, $4$, and $2$
dimensions, we have respectively:

\begin{eqnarray}
\bar{{\cal D}}_{5}^{\it (int)}(X)\ =\ e^{-0.518 X^{1/\sigma}}\ \times
\nonumber \\
(0.112 + 0.459 X - 0.260 X^2 + 0.201 X^3 - 0.050 X^4)
\label{eq:aval_fit_5d}
\end{eqnarray}
\begin{eqnarray}
\bar{{\cal D}}_{4}^{\it (int)}(X)\ =\ e^{-0.954 X^{1/\sigma}}\ \times
\nonumber \\
(0.058 + 0.396 X + 0.248 X^2 - 0.140 X^3 + 0.026 X^4)
\label{eq:aval_fit_4d}
\end{eqnarray}
\begin{eqnarray}
\bar{{\cal D}}_{2}^{\it (int)}(X)\ =\ e^{-1.076 X^{1/\sigma}}\ \times
\nonumber \\
(0.492 - 4.472 X + 14.702 X^2 - \nonumber \\
20.936 X^3 + 11.303 X^4)
\label{eq:aval_fit_2d}
\end{eqnarray}
with $1/\sigma=2.35, 3.20$, and $10.0$. The errors in the fits are again in
the 10\% range, judging from mean--field theory~\cite{OlgaLong}.

In mean field theory (dimensions six and greater) a similar fit~\cite{OlgaLong}
to the analytical form of the scaling function above $R_c$ gives
\begin{eqnarray}
\bar{{\cal D}}_{\rm MF}^{\it (int)}(X)\ =\ e^{-{X^2 \over 2}}\ &
(0.204 + 0.482 X - 0.391 X^2 + \nonumber \\
 & 0.204 X^3 - 0.048 X^4)
\label{eq:int_aval3a}
\end{eqnarray}

It is clear from the figure that the growing bump in the scaling curves
as the dimension decreases is a foreshadowing of a zero in the scaling
curve in two dimensions: this will be discussed further in
reference~\cite{Kuntz}.

\subsubsection{Binned in $H$ Avalanche Size Distribution}

The avalanche size distribution can also be measured at a field $H$ or
in a small range of fields centered around $H$. We have measured this
${\it binned}$ in $H$ avalanche size distribution for systems at the
critical disorder $R_c$ ($r=0$). To obtain the scaling form, we start
from the distribution of avalanches at field $H$ and disorder $R$
\begin{equation}
D(S,R,H) \sim S^{-\tau}\ {\cal D}_{\pm}(S^\sigma |r'|, |h|/|r'|^{\beta\delta})
\label{eq:aval_distr1}
\end{equation}
where as before ${\cal D}_{\pm}$ is the scaling function and $\pm$
indicates the sign of $r$. 

The parameter $B$ of equation~\ref{eq:scaling_2}, which rotates the measured
axes $(r,h)$ into the scaling axes $(r, h'=h+B r)$, will be
important~\cite{OlgaLong} only for
large avalanches of size $S > h^{-1/\sigma}$ near the critical point. In
three and four dimensions, this does not affect our scaling collapses; in
five dimensions we account for it~\cite{OlgaLong}.

The scaling function can be rewritten as ${\hat {\cal
D}}_{\pm}\Bigl(S^\sigma |r|, (S^\sigma |r|)^{\beta\delta}
|h|/|r|^{\beta\delta}\Bigr)$, where ${\hat D}_{\pm}$ is a new scaling
function and $\pm$ represents whether $H$ is greater than or less than $H_c$
({\it i.e.}, $H$ rather than $R$). Letting $R \rightarrow R_c$, the scaling
for the avalanche size distribution at the field $H$, measured at the
critical disorder $R_c$ is:
\begin{equation}
D(S,H) \sim S^{-\tau}\ {\hat {\cal D}}_{\pm}(|h| S^{\sigma\beta\delta})
\label{eq:aval_distr2}
\end{equation}

\begin{figure}
  \begin{center}
    \epsfxsize=8cm
    \epsffile{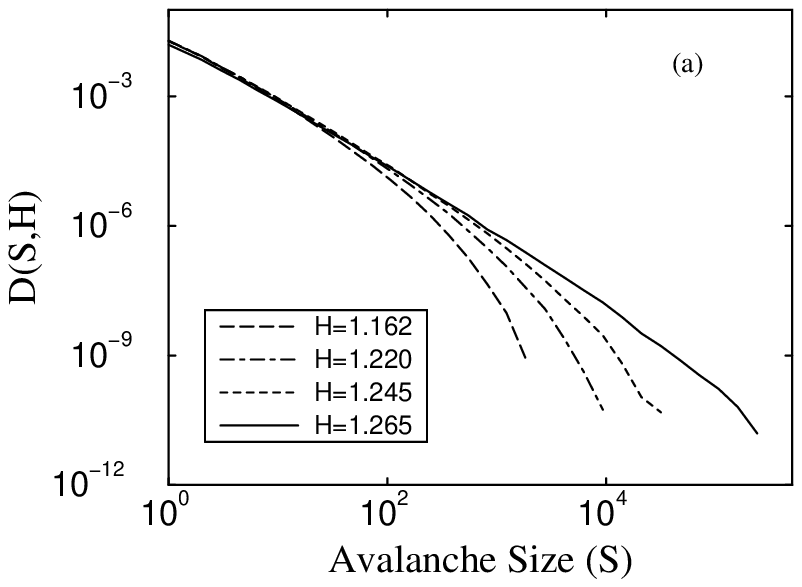}
    \epsfxsize=8cm
    \epsffile{
	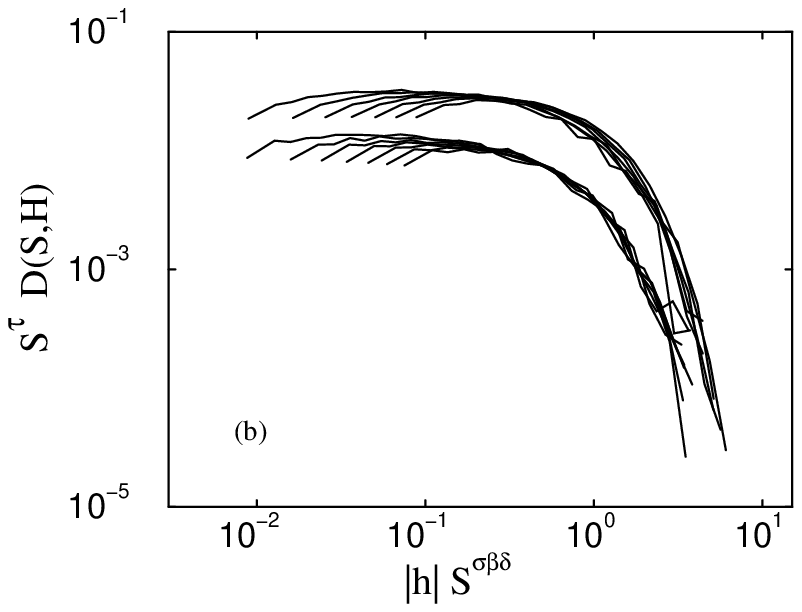}
  \end{center}
\caption{(a) {\bf Binned in $H$ avalanche size distribution in $4$
dimensions} for a system of $80^4$ spins at $R=4.09$ ($R_c=4.10$). The
critical field is $H_c=1.265$. The curves are averages over close to
$60$ random field configuration. Only a few curves are shown. (b)
Scaling collapse of the binned avalanche size distribution for $H<H_c$
(upper collapse) and $H>H_c$ (lower collapse). The critical exponents
are $\tau=1.53$ and $\sigma\beta\delta=0.54$, and the critical field is
$H_c=1.265$. The bins are at fields $1.162$, $1.185$, $1.204$, $1.220$,
$1.234$, $1.245$, $1.254$, $1.276$, $1.285$, $1.296$, $1.310$, $1.326$,
$1.345$, and $1.368$.}
\label{fig:bin_aval_4d}
\end{figure}

Figure~\ref{fig:bin_aval_4d}a shows the binned in $H$ avalanche size
distribution curves in $4$ dimensions, for values of $H$ below the
critical field $H_c$. (The curves and analysis are similar in $3$ and
$5$ dimensions; results in $4$ dimensions are used here for variety.)
The simulation was done at the best estimate of the critical disorder
$R_c$ ($4.1$ in $4$ dimensions). The binning in $H$ is logarithmic and
started from an approximate critical field $H_c$ obtained from the
magnetization curves; better estimates of $H_c$ are then obtained from the
binned distribution data curves and their collapses. Our best estimate
for the critical field $H_c$ in $4$ dimensions is $1.265 \pm 0.007$. The
scaling form for the logarithmically binned data is the same as in
equation~(\ref{eq:aval_distr2}), if the log-binned data is normalized by
the size of the bin. Figure~\ref{fig:bin_aval_4d}b shows the scaling
collapse for our data, both below and above the critical field
$H_c$. The ``top'' collapse gives the shape of the ${\hat {\cal D}_{-}}$
($H<H_c$) function, while the ``bottom'' collapse gives the ${\hat {\cal
D}_{+}}$ ($H>H_c$) function. Above the critical field $H_c$, there are
spanning avalanches in the system~\cite{note4}. These are not included
in the binned avalanche size distribution collapse shown in
figure~\ref{fig:bin_aval_4d}b.

The exponent $\tau$ which gives the power law behavior of the binned
avalanche size distribution is obtained from collapses of neighboring curves
as described above~\cite{OlgaLong}, extrapolating to $H=H_c$.
The exponent $\sigma\beta\delta$ is found
to be very sensitive to $H_c$, while $\tau$ is not. We have therefore
used the values of $\tau + \sigma\beta\delta$ and $\sigma$ from the
integrated avalanche size distribution collapses, and $\tau$ from the
binned avalanche size distribution collapses to further constrain $H_c$
(by constraining $\sigma\beta\delta$), and to calculate $\beta\delta$.
The latter is then used to obtain collapses of the magnetization curves.
We should mention here that $H_c$ in all the dimensions is difficult to
find and that it is influenced by finite sizes. The values listed in
Table~\ref{table:RH} are the best estimates obtained from the largest
system sizes we have. Nevertheless, systematic errors for $H_c$ could be
larger than the errors given in Table~\ref{table:RH}. These errors 
could produce systematic errors for $\sigma\beta\delta$ which depends on
$H_c$, and for $\beta\delta$ which is calculated from
$\sigma\beta\delta$: hence errors in these exponents could also be larger
than the errors listed in Table~\ref{table:calculated_exp}.

\begin{figure}
  \begin{center}
    \epsfxsize=8cm
    \epsffile{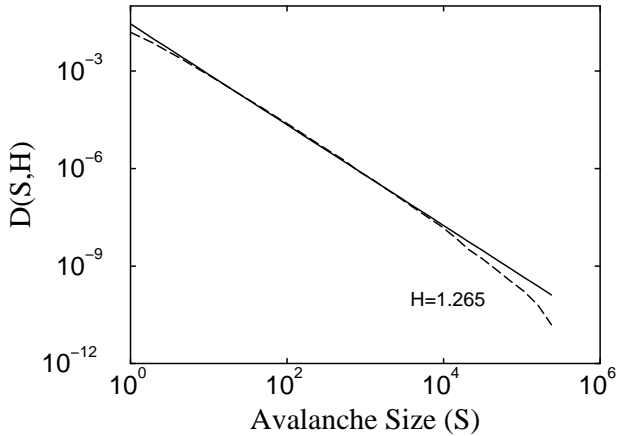}
  \end{center}
\caption{{\bf Linear fit to binned avalanche size distribution curve in
$4$ dimensions}, for a system of $80^4$ spins at $R_c=4.09$. The
magnetic field is $H=1.265$. The straight solid line is a linear fit to
the data for $S < 13,000$ spins. The slope from the fit is $1.55$ (this
varies by not more than $3\%$ as the range over which the fit is done is
changed), while the exponent $\tau$ obtained from the collapses and the
extrapolation in figure~\protect\ref{fig:bin_aval_4d} is $1.53 \pm
0.08$.}
\label{fig:bin_aval_fit_4d}
\end{figure}

From figure~\ref{fig:bin_aval_4d}b, we see that the two binned avalanche
size distribution scaling function do not have a ``bump'' as does the
scaling function for the integrated avalanche size distribution (inset
in figure~\ref{fig:aval_3}). Therefore, we expect that the exponent
$\tau$ which gives the slope of the distribution in
figure~\ref{fig:bin_aval_4d}a can also be obtained by a linear fit through the
data curve closest to the critical field. Figure~\ref{fig:bin_aval_fit_4d}
shows the curve for the $H=1.265$ bin (dashed
curve) as well as the linear fit. The slope from the linear fit is
$1.55$ while the value of $\tau$ obtained from the collapses and the
extrapolation in figure~\ref{fig:bin_aval_4d} is $1.53 \pm 0.08$.

\subsection{Avalanche Correlations}

\begin{figure}
  \begin{center}
    \epsfxsize=8cm
%    \epsffile{Figures/Norm_correl_d3_L320_paper_THESIS.ps}
    \epsffile{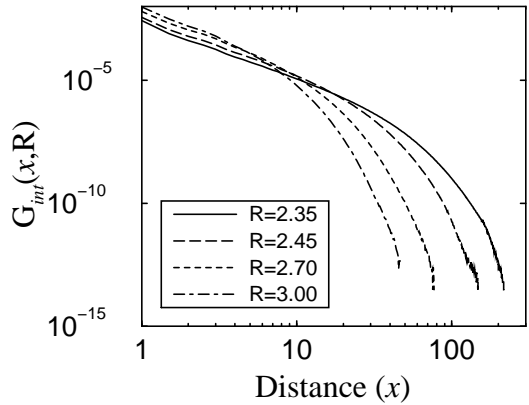}
  \end{center}
\caption{{\bf Avalanche correlation function integrated over the field
$H$ in $3$ dimensions}, for $L=320$. The curves are averages of up to
$19$ random field configurations. The critical disorder $R_c$ is $2.16$.
}
\label{fig:correl_3d}
\end{figure}

The avalanche correlation function $G(x,R,H)$ measures the probability
that the initial spin of an avalanche will trigger, in that avalanche,
another spin a distance $x$ away. From the renormalization group
description\cite{Dahmen1,Dahmen2}, close to the critical point and for
large distances $x$, the correlation function is given by:
\begin{equation}
G(x,R,H) \sim {1 \over {x^{d-2+\eta}}}\ {\cal G}_{\pm}(x/{\xi(r,h)})
\label{eq:correl_1}
\end{equation}
where $r$ and $h$ are respectively the reduced disorder and field,
${\cal G_{\pm}}$ ($\pm$ indicates the sign of $r$) is the scaling
function, $d$ is the dimension, $\xi$ is the correlation length, and
$\eta$ is called the ``anomalous dimension''. Corrections can be shown
to be subdominant~\cite{OlgaLong}. The correlation length
$\xi (r,h)$ is a macroscopic length scale in the system which is 
on the order of the mean linear extent of the largest avalanches. 
At the critical field $H_c$ (h=0) and near
$R_c$, the correlation length scales like $\xi \sim |r|^{-\nu}$, while
for small field $h$ it is given by 
\begin{equation}
\xi \sim |r|^{-\nu}\ {\cal Y}_{\pm}(h/|r|^{\beta\delta})
\label{eq:correl_1b}
\end{equation}
where ${\cal Y}_{\pm}$ is a universal scaling function. The avalanche
correlation function should not be confused with the cluster or
``spin-spin'' correlation which measures the probability that two spins
a distance $x$ away have the same value. (The algebraic decay for this
other, spin-spin correlation function at the critical point ($r=0$ and
$h=0$), is $1/{x^{d-4+{\tilde \eta}}}$\cite{Dahmen1}.)

We've mostly used, for historical reasons, a slightly different
avalanche correlation function, which scales the same way as the
``triggered'' correlation function $G$ described above. Our function
basically ignores the difference between the triggering spin and the
other spins in the avalanche: alternatively, it calculates for
avalanches of size $S$ the correlation function for pairs of spins, and
then averages over all avalanches (weighting each avalanche equally).
We've checked that our correlation function agrees to within 3\% with
the ``triggered'' correlation function described above, for $R>R_c$ in
three dimensions and above. (In two dimensions, the two definitions
differ more substantially, but appear to scale in the same
way~\cite{Kuntz}.)

We have measured the avalanche correlation function integrated over the
field $H$, for $R>R_c$. For every avalanche that occurs between
$H=-\infty$ and $H =+\infty$, we keep a count on the number of times a
distance $x$ occurs in the avalanche. To decrease the computational time
not every pair of spins is selected; instead we do a statistical
sampling~\cite{Numerical}. The spanning avalanches are not included in
our correlation measurement. Figure~\ref{fig:correl_3d} shows several
avalanche correlation curves in $3$ dimensions for $L=320$. The scaling
form for the avalanche correlation function integrated over the field
$H$, close to the critical point and for large distances $x$, is
obtained by integrating equation (\ref{eq:correl_1}):
\begin{equation}
G_{\it int}(x,R) \sim \int {1 \over {x^{d-2+\eta}}}\
{\cal G}_{\pm}\Bigl(x/{\xi(r,h)}\Bigr)\ dh
\label{eq:correl_2}
\end{equation}
Using equation (\ref{eq:correl_1b}) and defining $u=h/|r|^{\beta\delta}$,
equation (\ref{eq:correl_2}) becomes:
\begin{equation}
G_{\it int}(x,R) \sim |r|^{\beta\delta}{x^{-(d-2+\eta)}} \int
{\cal G}_{\pm}\Bigl(x/|r|^{-\nu} {\cal Y}_{\pm} (u)\Bigr)\ du
\label{eq:correl_3}
\end{equation}
The integral ($\cal I$) in equation (\ref{eq:correl_3}) is a function of
$x|r|^{\nu}$ and can be written as:
\begin{equation}
{\cal I} =
(x|r|^{\nu})^{-\beta\delta/\nu}\ {\widetilde {\cal G}}_{\pm}(x|r|^{\nu})
\label{eq:correl_4}
\end{equation}
to obtain the scaling form:
\begin{equation}
G_{\it int}(x,R) \sim {1 \over x^{d+\beta/\nu}}\
{\widetilde {\cal G}}_{\pm}(x|r|^{\nu})
\label{eq:correl_5}\end{equation}
where we have used the scaling relation $(2-\eta)\nu=\beta\delta-\beta$
(see~\cite{Dahmen1} for the derivation).

\begin{figure}
  \begin{center}
    \epsfxsize=8cm
    \epsffile{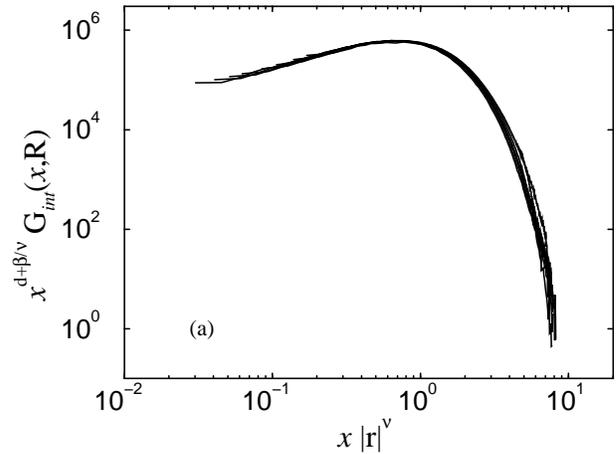}
  \end{center}
\caption{{\bf Scaling collapse of the avalanche correlation function
integrated over the field $H$, in $3$ dimensions} for $L=320$. The
values of the disorders range from $R=2.35$ to $R=3.0$, with $R_c=2.16$.
The exponents used in the collapse are $\nu=1.39 \pm 0.20$ and $d +
\beta/\nu = 3.07 \pm 0.30$. When collapses of neighboring curves are
extrapolated to $R_c$, we get a slightly smaller value of $\nu= 1.37 \pm
0.18$.}
\label{fig:correl_collapse_3d}
\end{figure}

\begin{figure}
  \begin{center}
    \epsfxsize=8cm
    \epsffile{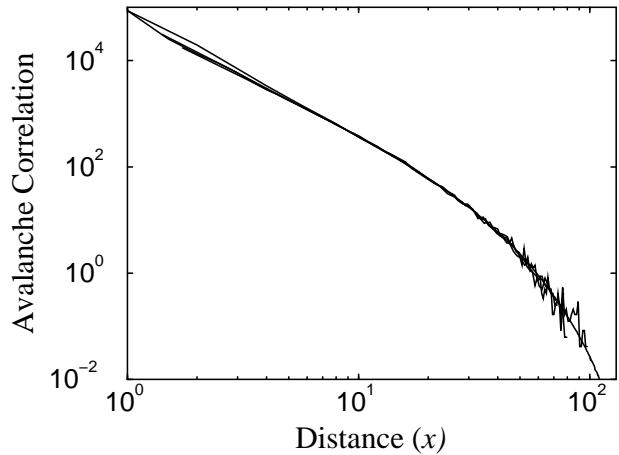}
  \end{center}
\caption{{\bf Anisotropies in the avalanche correlation function}. The
curves are for a system of $320^3$ spins at $R=2.35$. Four curves are
shown on the graph: one is the avalanche correlation function integrated
over the field $H$ (as in figure~\protect\ref{fig:correl_3d}), while the
other three are measurements of the correlation along the three axis,
the six face diagonals, and the four body diagonals. Avalanches
involving more than four spins show no noticeable anisotropy: the
critical point appears to have spherical symmetry. The same result is
found in $2$ dimensions.}
\label{fig:correl_aniso_3d}
\end{figure}

Figure~\ref{fig:correl_collapse_3d} shows the integrated avalanche
correlation curves collapse in $3$ dimensions for $L=320$ and $R>R_c$.
The exponent $\nu$ is obtained from such collapses by extrapolating to
$R = R_c$ as was done for other collapses~\cite{OlgaLong}. The exponent
$\beta/\nu$ can be obtained from these collapses too, but it is much
better estimated from the magnetization discontinuity covered below. The
value of $\beta/\nu$ listed in Table~\ref{table:exponents} is derived
exclusively from the magnetization discontinuity collapses.

We have also looked for possible anisotropies in the integrated
avalanche correlation function in $2$ and $3$ dimensions. The
anisotropic integrated avalanche correlation functions are measured
along ``generalized diagonals'': one along the three axis, the second
along the six face diagonals, and the third along the four body
diagonals. We compare the integrated avalanche correlation function and
the anisotropic integrated avalanche correlation functions to each
other, and find no anisotropies in the correlation, as can be seen from
figure~\ref{fig:correl_aniso_3d}.

\subsection{Spanning Avalanches}

\begin{figure}
  \begin{center}
    \epsfxsize=7.cm
    \epsffile{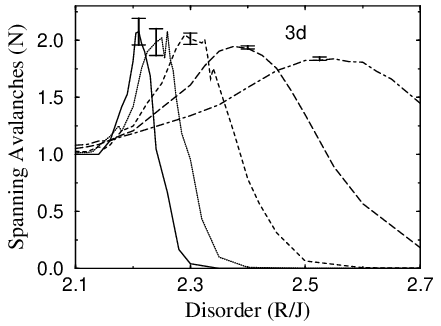}
    \epsfxsize=7.cm
    \epsffile{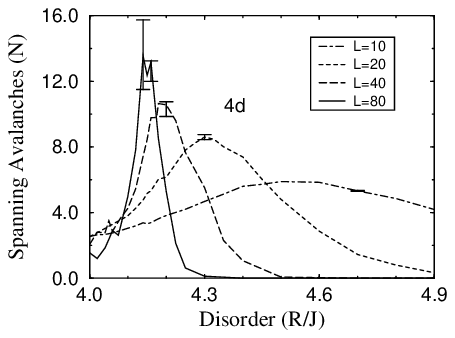}
    \epsfxsize=7.cm
    \epsffile{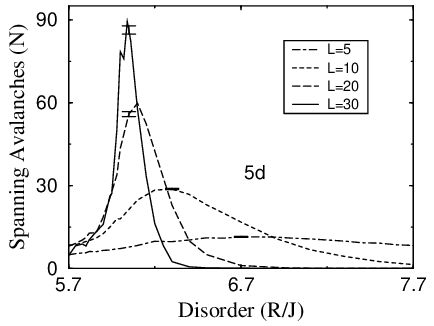}
    \epsfxsize=7.cm
    \epsffile{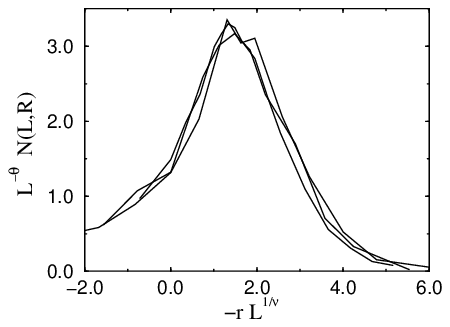}
  \end{center}
\caption{{\bf Spanning avalanches in $3$, $4$, and $5$ dimensions.} (a)
{\bf Number of spanning avalanches $N$ in $3$ dimensions,} occurring in
the system between $H= -\infty$ to $H=\infty$, as a function of the
disorder $R$, for linear sizes $L$: $20$ (dot-dashed), $40$ (long
dashed), $80$ (dashed), $160$ (dotted), and $320$ (solid). The critical
disorder $R_c$ is at $2.16$. The error bars for each curve tend to be
smaller than the error bar shown at the peak for disorders above the
peak and larger for disorders below the peak. They are not given here
for clarity. Note that the number of avalanches increases only slightly
as the size is increased. (b) {\bf Number of spanning avalanches in $4$
dimensions.} The critical disorder is $4.1$. (c) {\bf Number of spanning
avalanches in $5$ dimensions.} The critical disorder is $5.96$. Both in
$4$ and $5$ dimensions, the peaks grow and shift towards $R_c$ as the
size of the system is increased. (d) {\bf Collapse of the spanning avalanche
curves in $4$ dimensions} for linear sizes $L=20,40$, and $80$. The
exponents are $\theta = 0.32$ and $\nu = 0.89$, and the critical
disorder is $R_c = 4.10$. The collapse is done using $r = (R_c-R)/R$.}
\label{fig:span_aval_345}
\end{figure}

The critical disorder $R_c$ was defined earlier as the disorder $R$ at
which an ${\it infinite}$ avalanche first appears in the system, in the
thermodynamic limit, as the disorder is lowered. At that point, the
magnetization curve will show a discontinuity at the magnetization
$M_c(R_c)$ and field $H_c(R_c)$. For each disorder $R$ below the
critical disorder, there is ${\it one}$ infinite avalanche that occurs
at a critical field $H_c(R)>H_c(R_c)$~\cite{Dahmen1,Dahmen2}, while
above $R_c$ there are only finite avalanches. This is the behavior for
an infinite size system. In a finite size system far below and above
$R_c$ the above picture is still true, but close to the critical
disorder, as we approach the transition, the avalanches get larger and
larger, and there will be a first point where one of them will span the
system from one side to another in at least one direction. This
avalanche is not the infinite avalanche; if the system was larger, this
avalanche would typically be non--system spanning. Such an avalanche
(which spans the system) we call a spanning avalanche.

In our numerical simulation, we find that for finite sizes $L$, there
are not one but ${\it many}$ such avalanches in $4$ and $5$ dimensions
(and maybe $3$), and that their number increases as the system size
increases~\cite{comment}. Figures~\ref{fig:span_aval_345}(a-c) show the
number of spanning avalanches as a function of disorder $R$, for
different sizes and dimensions. In $4$ and $5$ dimensions, the spanning
avalanche curves become more narrow as the system size is increased.
Also, the peaks shift toward the critical value of the disorder ($4.1$
and $5.96$ respectively), and the number of spanning avalanches at $R_c$
increases. This suggests that in $4$ and $5$ dimensions, for $L
\rightarrow \infty$, there will be one infinite avalanche below $R_c$,
none above, and an infinite number of infinite, spanning avalanches at
the critical disorder $R_c$. In $3$ dimensions, the results are not
conclusive, as noted both from figure~\ref{fig:span_aval_345}a and from
the value of the spanning avalanche exponent $\theta = 0.15 \pm 0.15$
defined below: a value of $\theta=0$ is consistent with one infinite or
spanning avalanche at $R_c$ as $L \rightarrow \infty$. It is clear that
$\theta=0$ in two dimensions, since spanning avalanches can't
interpenetrate: it's thus plausible that $\theta$ is near zero in three
dimensions because it must vanish one dimension lower.

In percolation, a similar multiplicity of infinite
clusters~\cite{Arcangelis,Stauffer} as the system size is increased is found
for dimensions above six which is the upper critical dimension (UCD).
The UCD is the dimension at and above which the mean field exponents are
valid. Below six dimensions, there is only one such infinite cluster in
percolation.
The existence of a diverging number of infinite clusters in percolation
is associated with the breakdown of the hyperscaling relation above six
dimensions. Since a hyperscaling relation is a relation between critical
exponents that includes the dimension $d$ of the system, it is always
only satisfied up to and including the upper critical dimension. In our
system, the upper critical dimension is also six, but we find spanning
avalanches in dimensions even below that. In a comment by Maritan {\it
{et al.}}\cite{comment}, it was suggested that our system should satisfy
the hyperscaling relation $d\nu-\beta = 1/\sigma$ 
found in percolation~\cite{Stauffer}. But since our system has spanning
avalanches below the upper critical dimension, this hyperscaling
relation breaks down below six dimensions. Due to the existence of many
spanning avalanches near $R_c$, the new ``violation of hyperscaling''
relation for dimensions three and above becomes~\cite{Dahmen1}:
\begin{equation}
(d-\theta)\nu - \beta = 1/\sigma
\label{eq:span_aval_1}
\end{equation}
where $\theta$ is the ``breakdown of hyperscaling'' or spanning
avalanches exponent defined below. One can check that our exponents in
$3$, $4$, and $5$ dimensions and mean field satisfy this equation (see
Tables~\ref{table:exponents} and~\ref{table:calculated_exp}).

For the simulation, we define a spanning avalanche to be an avalanche
that spans the system in a particular direction. We average over all the
directions to obtain better statistics. Depending on the size and
dimension of the system and the distance from the critical disorder, the
number of spanning avalanches for a particular value of disorder $R$ is
obtained by averaging over as few as $5$ to as many as $2000$ different
random field configurations. We define the exponent $\theta$ such that
the number $N$ of spanning avalanches, at the critical disorder $R_c$,
increases with the linear system size as: $N \sim L^{\theta}$ ($\theta >
0$). The finite size scaling form\cite{RG_FiniteSize} for the number of
spanning avalanches close to the critical disorder is:
\begin{equation}
N(L,R) \sim L^{\theta}\ {\cal N}_{\pm}(L^{1/\nu}|r|)
\label{eq:span_aval2}
\end{equation}
where $\nu$ is the correlation length exponent and ${\cal N}_{\pm}$ is
the corresponding scaling function ($\pm$ indicates the sign of $r$).
The collapse is shown in figure~\ref{fig:span_aval_345}d. (We
show the collapses in $4$ dimensions here since the
existence of spanning avalanches in $3$ dimensions is not conclusive.)
These values are used along with the results from other collapses to
obtain Table~\ref{table:exponents}. In the analysis of the avalanche
size distribution, magnetization, and correlation functions for $R>R_c$,
how close we chose to come to the critical disorder $R_c$ was determined
by the spanning avalanches: we include no values $R$ below the first
value which exhibited a spanning avalanche.

\subsection{Magnetization Discontinuity}

We have mentioned earlier that in the thermodynamic limit, at and below
the critical disorder $R_c$, there is a critical field $H_c(R)>H_c(R_c)$
at which the infinite avalanche occurs. Close to the critical
transition, for small $r<0$, the change in the
magnetization due to the infinite avalanche scales as (equation
(\ref{eq:scaling_1})):
\begin{equation}
\Delta M(R) \sim\ r^{\beta}
\label{eq:deltaM_1}
\end{equation}
where $r=(R_c-R)/R$, while above the transition, there is no infinite
avalanche.

In finite size systems, the transition is not as sharp: we
have spanning avalanches above the critical disorder. If we measure the
change in the magnetization due to all the spanning avalanches as
a function of disorder $R$ at various system sizes $L$, we expect it will
obey finite--size scaling (as did the number of spanning avalanches):
\begin{equation}
\Delta M(L,R) \sim\ |r|^{\beta}\ \Delta {\cal M}_{\pm}(L^{1/\nu}|r|)
\label{eq:deltaM_2}
\end{equation}
where $\Delta {\cal M}_{\pm}$ is a universal scaling function. (The
parameter $B$ here (equation~\ref{eq:scaling_2})
is unimportant~\cite{OlgaLong} because we $\Delta M$
is measured at $h'=0$.) Defining a new universal scaling
function $\Delta \widetilde {\cal M}_{\pm}$:
\begin{equation}
\Delta {\cal M}_{\pm}(L^{1/\nu}|r|) \equiv\ (L^{1/\nu}|r|)^{-\beta}\
\Delta \widetilde {\cal M}_{\pm}(L^{1/\nu}|r|)
\label{eq:deltaM_3}
\end{equation}
we obtain the scaling form:
\begin{equation}
\Delta M(L,R) \sim\ L^{-\beta/\nu}\ \Delta {\widetilde {\cal M}_{\pm}}
(L^{1/\nu}|r|)
\label{eq:deltaM_4}
\end{equation}

\begin{figure}
  \begin{center}
    \epsfxsize=8cm
    \epsffile{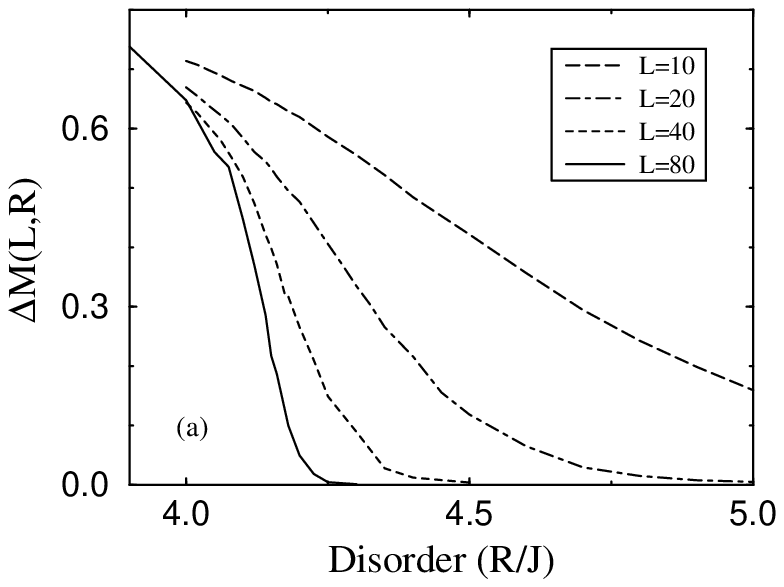}
    \epsfxsize=8cm
    \epsffile{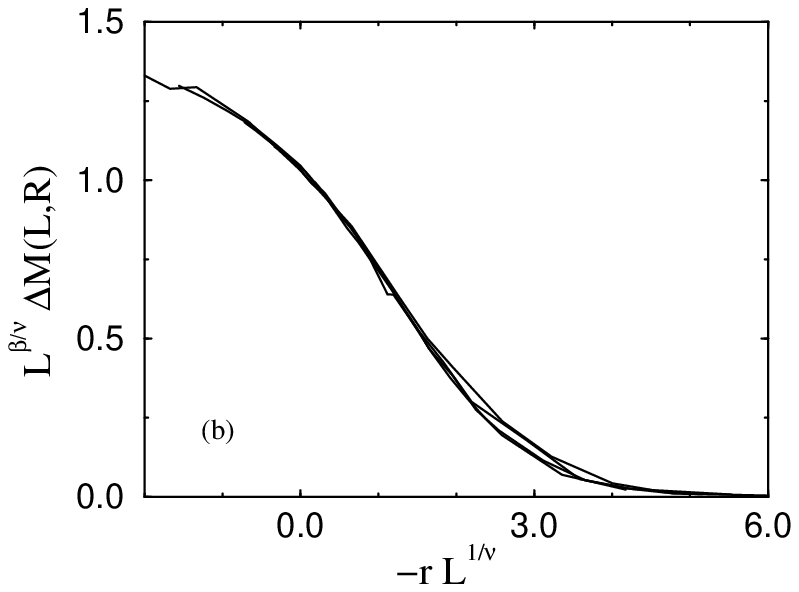}
  \end{center}
\caption{{\bf Jump in the magnetization, in $4$ dimensions}
(a)~Change in the magnetization due to the spanning avalanches in $4$
dimensions, for several linear sizes $L$, as a function of the disorder
$R$. (b) Scaling collapse of the curves in (a) using $r=(R_c-R)/R$. The
exponents are $1/\nu = 1.12$ and $\beta/\nu = 0.19$, and the critical
disorder is $R_c = 4.1$.}
\label{fig:deltaM_4d}
\end{figure}

Figures~\ref{fig:deltaM_4d}a and \ref{fig:deltaM_4d}b show the change
in the magnetization due to the spanning avalanches in $4$ dimensions,
and a scaling collapse of that data (similar results exist in $3$ and
$5$ dimensions). Notice that as the system size increases, the curves
approach the $|r|^\beta$ behavior. The exponents $1/\nu$ and
$\beta/\nu$ are extracted from scaling collapses (figure~\ref{fig:deltaM_4d}b)
and extrapolated to $R_c$~\cite{OlgaLong}.
The value of $\beta$ is calculated from $\beta/\nu$ and the knowledge of
$\nu$, and is the value used for collapses of the magnetization curves
(discussed earlier).

\subsection{Moments of the Avalanche Size Distribution}

\begin{figure}
  \begin{center}
    \epsfxsize=8cm
    \epsffile{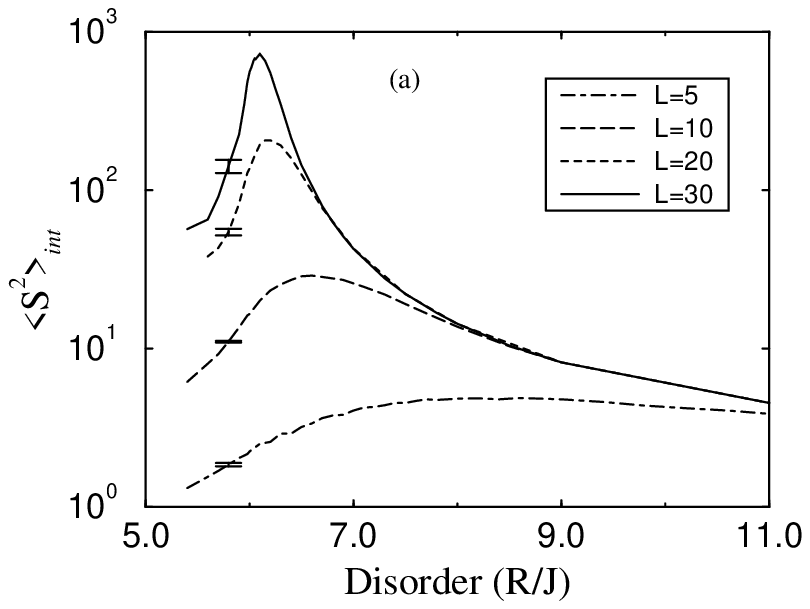}
    \epsfxsize=8cm
    \epsffile{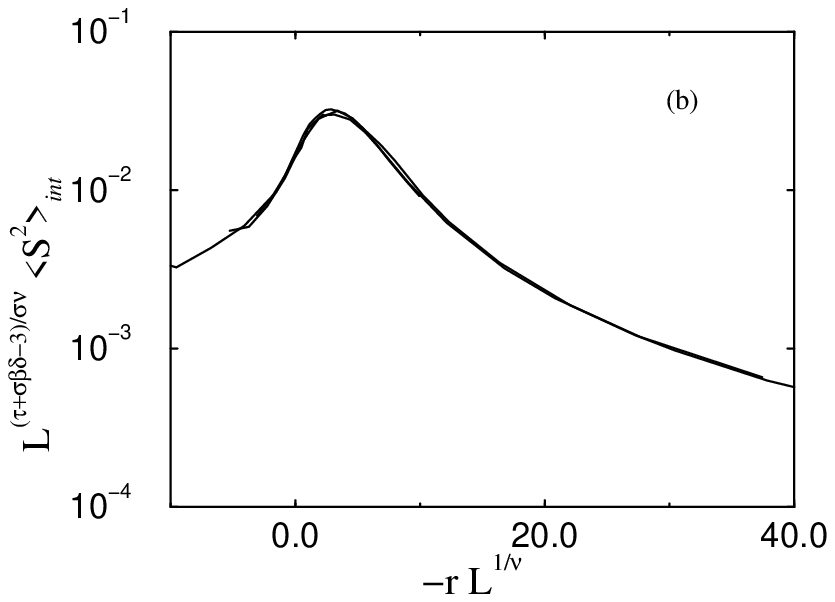}
  \end{center}
\caption{{\bf Second moments}. (a)~Second moments of the avalanche size
distribution integrated over the field $H$, in $5$ dimensions. Error
bars are largest for smaller disorders (shown on the curves). The curves
have between $24$ and $50$ points, and the value of the second moment
for each disorder is averaged over $3$ to $100$ different random field
configurations. (b)~Scaling collapse of the $L=10, 20$, and $30$ curves
from~(a) using $r=(R_c-R)/R$. The exponents are $1/\nu = 1.47$ and $\rho
= -(\tau+\sigma\beta\delta - 3)/\sigma\nu = 2.95$, and the critical
disorder is $R_c = 5.96$.}
\label{fig:s2_5d}
\end{figure}

The second moment of the integrated avalanche size distribution 
has a finite--size scaling form 
\begin{equation}
\langle S^2 {\rangle}_{\it int} \sim L^{-(\tau+\sigma\beta\delta
- 3)/\sigma\nu}\
{\widetilde {\cal S}}_{\pm}^{(2)}(L^{1/\nu}|r|)
\label{eq:s2_5d}
\end{equation}
where $L$ is the linear size of the system, $r$ is the reduced disorder,
$\widetilde {\cal S}_{\pm}^{(2)}$ is the scaling function, and $\nu$ is
the correlation length exponent. We can similarly define the third and
fourth moment, with the exponent $-(\tau+\sigma\beta\delta -
3)/\sigma\nu$ replaced by $-(\tau+\sigma\beta\delta-4)/\sigma\nu$ and
$-(\tau+\sigma\beta\delta -5)/\sigma\nu$ respectively.
Figures~\ref{fig:s2_5d}a and~\ref{fig:s2_5d}b show the second moments
data in $5$ dimensions for sizes $L=5, 10, 20,$ and $30$, and a collapse
(again, results in $3$ and $4$ dimensions are similar and we have chosen
to show the curves in $5$ dimensions for variety). The curves are
normalized by the average avalanche size integrated over all fields $H$:
$\int_{-\infty}^{+\infty} \int_{1}^{\infty} S\ D(S,R,H,L)\ dS\ dH$. The
spanning avalanches are not included in the calculation of the moments.
We omit the $L=5$ curve from the collapse; it doesn't collapse with the
others well, presumably because of subdominant finite size effects. The
exponents for the third and fourth moment can be calculated from those
of the second moment, and we find that they agree with the values
obtained from their respective collapses.

\subsection{Avalanche Time Measurement}

The exponents we have measured so far are static scaling exponents: they
do not depend on the dynamics of the model. If we measure the time an
avalanche takes to occur, we are making a dynamical measurement. The
time measurement in the numerical simulation is done by increasing the
time clock by one for each shell of spins in the avalanche. That is,
we implement time as a synchronous dynamics, where in each time step
all unstable spins from the previous step are flipped.
The scaling relation between the time $t$ it takes
an avalanche to occur and the linear size $\xi$ of the avalanche 
defines the critical exponent $z$~\cite{dynamicz,MaBinney}:
\begin{equation}
t \sim \xi^z
\label{eq:time_1}
\end{equation}
The exponent $z$ is known as the dynamical critical exponent. 
Equation~(\ref{eq:time_1}) gives the scaling for the time it takes for a
spin to ``feel'' the effect of another a distance $\xi$ away. Since the
correlation length $\xi$ scales like $r^{-\nu}$ close to the critical
disorder, and the characteristic size $S$ as $r^{-1/\sigma}$, the time
$t$ then scales with avalanche size as:
\begin{equation}
t \sim S^{\sigma \nu z}
\label{eq:time_2}
\end{equation}

In our simulation, we measure the distribution of times for each
avalanche size $S$. The distribution of times $D_t(S,R,H,t)$ for an
avalanche of size $S$ close to the critical field $H_c$ and critical
disorder $R_c$ is
\begin{equation}
D_t(S,R,H,t) \sim S^{-q}\ {\bar {\cal D}}_{\pm}^{(t)} (S^{\sigma}|r|,
h/|r|^{\beta\delta}, t/S^{\sigma\nu z})
\label{eq:time_3}
\end{equation}
where $q=\tau +\sigma\nu z$, and is defined such that
\begin{eqnarray}
\int_{-\infty}^{+\infty} \!\! \int_{1}^{\infty} D_t(S,R,H,t)\ dH\ dt\ =\
\nonumber \\
S^{-(\tau + \sigma\beta\delta)}\ {\bar {\cal D}}_{\pm}^{(int)}(S^{\sigma}|r|)
\label{eq:time_4}
\end{eqnarray}
where ${\bar {\cal D}}_{\pm}^{(int)}$ was defined in the integrated
avalanche size distribution section. The avalanche time distribution
integrated over the field $H$, at the critical disorder ($r=0$) is:
\begin{equation}
D_t^{(int)}(S,t)\ \sim\
t^{-{(\tau + \sigma\beta\delta + \sigma\nu z) /\sigma\nu z}}\ 
{\cal D}_t^{(int)}(t/S^{\sigma\nu z})
\label{eq:time_5}
\end{equation}
which obtained from equation (\ref{eq:time_3}) (reference~\cite{OlgaLong}).

\begin{figure}
  \begin{center}
    \epsfxsize=8cm
    \epsffile{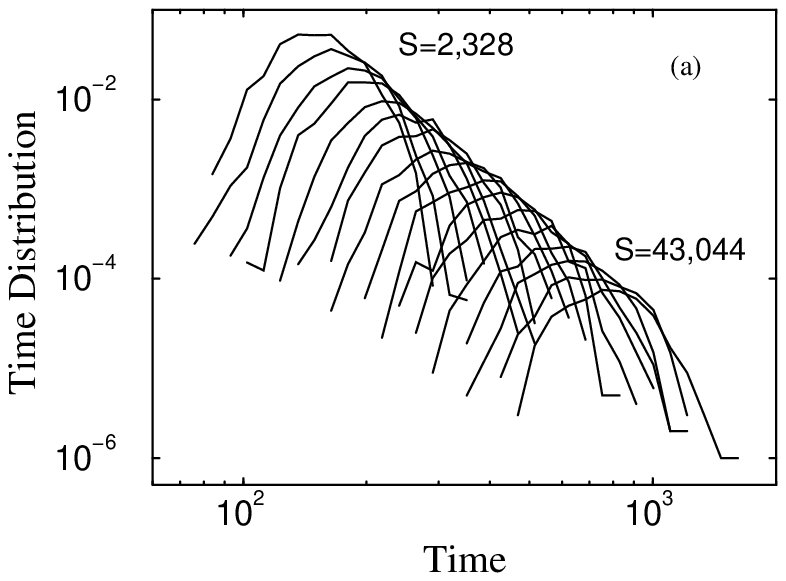}
    \epsfxsize=8cm
    \epsffile{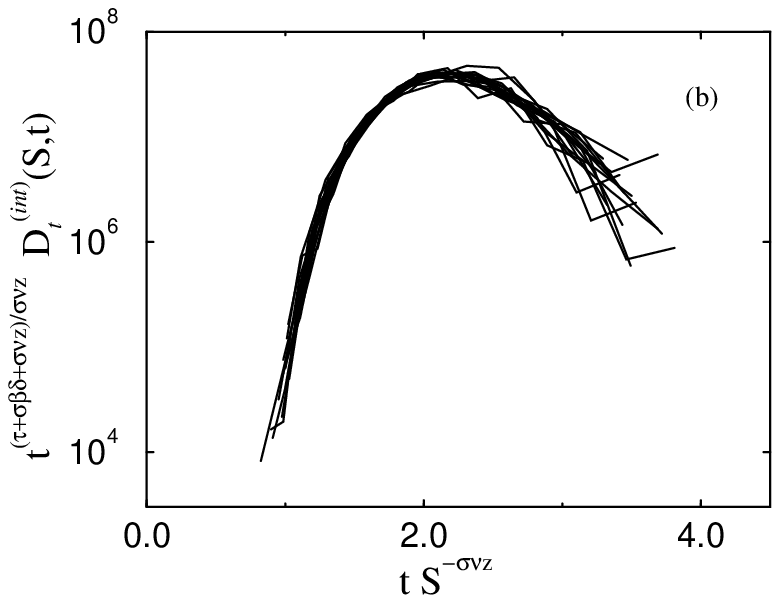}
  \end{center}
\caption{(a)~{\bf Avalanche time distribution curves in $3$ dimensions,}
for avalanche size bins from about $2000$ to $40000$ spins (from upper
left to lower right corner). The system size is $800^3$ at $R=2.26$. The
curves are from only one random field configuration. (b)~Scaling
collapse of curves in~(a). The values of the exponents are $\sigma\nu z
= 0.57$ and $(\tau+\sigma\beta\delta+ \sigma\nu z)/\sigma\nu z = 4.0$.}
\label{fig:time_3d}
\end{figure}

Figures~\ref{fig:time_3d}a and~\ref{fig:time_3d}b show the avalanche
time distribution integrated over the field $H$ for different avalanche
sizes, and a collapse of these curves using the above scaling form, for
a $800^3$ system at $R=2.260$ (just above the range where spanning
avalanches occur). The data is saved in logarithmic size bins, each
about $1.2$ times larger than the previous one. The time is also
measured logarithmically (next bin is $1.1$ times larger than the
previous one). The extracted value for $z$ in $3$ dimensions is $1.68
\pm 0.07$. The results for other dimensions are listed in
Table~\ref{table:exponents}.

\onecolumn

\subsection{Tables of Results}

\begin{table}
%\narrowtext
\begin{tabular}{cr@{$\,\pm\,$}lr@{$\,\pm\,$}lr@{$\,\pm\,$}lc}
measured exponents&
\multicolumn{2}{c}{$3$d} &
\multicolumn{2}{c}{$4$d} &
\multicolumn{2}{c}{$5$d} &
\multicolumn{1}{c}{mean field}  \\ \hline
$1/\nu$ & 0.71 & 0.09 & 1.12 & 0.11 & 1.47 & 0.15 & 2 \\
$\theta$ & 0.015 & 0.015 & 0.32 & 0.06 & 1.03 & 0.10 & 1 \\
$(\tau+\sigma\beta\delta -3)/\sigma \nu$ & -2.90 & 0.16
& -3.20 & 0.24 & -2.95 & 0.13 & -3 \\
$1/\sigma$ & 4.2 & 0.3 & 3.20 & 0.25 & 2.35 & 0.25 & 2 \\
$\tau+\sigma\beta\delta$ & 2.03 & 0.03 & 2.07 & 0.03 & 2.15 & 0.04 & 9/4 \\
$\tau$ & 1.60 & 0.06 & 1.53 & 0.08 & 1.48 & 0.10 & 3/2 \\
$d + \beta/\nu$ & 3.07 & 0.30 & 4.15 & 0.20 & 5.1 & 0.4 & 7 (at
$d_c=6$)\\
$\beta/\nu$ & 0.025 & 0.020 & 0.19 & 0.05 & 0.37 & 0.08 & 1 \\
$\sigma\nu z$ & 0.57 & 0.03 & 0.56 & 0.03 & 0.545 & 0.025  & 1/2 \\
\end{tabular}
\vspace{0.25cm}
\caption{{\bf Universal critical exponents.}
Values for the exponents extracted from scaling collapses in $3$, $4$,
and $5$ dimensions. The mean field values are calculated
analytically\protect\cite{prl1,Dahmen1}. $\nu$ is the correlation
length exponent and is found from collapses of avalanche correlations,
number of spanning avalanches, and moments of the avalanche size
distribution data. The exponent $\theta$ is a measure of the number of
spanning avalanches and is obtained from collapses of that data.
$(\tau+\sigma\beta\delta-3)/\sigma\nu$ is obtained from the second
moments of the avalanche size distribution collapses. $1/\sigma$ is
associated with the cutoff in the power law distribution of avalanche
sizes integrated over the field $H$, while $\tau+\sigma\beta\delta$
gives the slope of that distribution. $\tau$ is obtained from the binned
avalanche size distribution collapses. $d+\beta/\nu$ is obtained from
avalanche correlation collapses and $\beta/\nu$ from magnetization
discontinuity collapses. $\sigma\nu z$ is the exponent combination for
the time distribution of avalanche sizes and is extracted from that
data. Error bars are based on variations in the results based on different
approaches to the analysis: statistical fluctuations are typically smaller.
}
\label{table:exponents}
\end{table}

\begin{table}
\begin{tabular}{cr@{$\,\pm\,$}lr@{$\,\pm\,$}lr@{$\,\pm\,$}lc}
calculated exponents &
\multicolumn{2}{c}{$3$d} &
\multicolumn{2}{c}{$4$d} &
\multicolumn{2}{c}{$5$d} &
\multicolumn{1}{c}{mean field}  \\ \hline
$\sigma \beta \delta$ & 0.43 & 0.07 & 0.54 & 0.08 & 0.67 & 0.11 &
3/4 \\
$\beta\delta$ & 1.81 & 0.32 & 1.73 & 0.29 & 1.57 & 0.31 & 3/2 \\
$\beta$ & 0.035 & 0.028 & 0.169 & 0.048 & 0.252 & 0.060 & 1/2 \\$\sigma\nu$ & 0.34 & 0.05 & 0.28 & 0.04 & 0.29 & 0.04 & 1/4 \\
$\eta = 2 + (\beta-\beta\delta)/\nu$ & 0.73 & 0.28 & 0.25 & 0.38 &
0.06 & 0.51 & 0 \\
\end{tabular}
\vspace{0.25cm}
\caption{Values for exponents in $3$, $4$, and $5$ dimensions that are
not extracted directly from scaling collapses, but instead are derived
from Table~\protect\ref{table:exponents} and the exponent relations
(see~\protect\cite{Dahmen1}). The mean field values are obtained
analytically~\protect\cite{prl1,Dahmen1}. Both $\sigma\beta\delta$ and
$\beta\delta$ could have larger systematic errors than the errors listed
here. See the binned avalanche size distribution section for details.}
\label{table:calculated_exp}
\end{table}

\begin{table}
%\narrowtext
\begin{tabular}{cr@{$\,\pm\,$}lr@{$\,\pm\,$}lr@{$\,\pm\,$}lc}
 &
\multicolumn{2}{c}{$3$d} &
\multicolumn{2}{c}{$4$d} &
\multicolumn{2}{c}{$5$d} &
\multicolumn{1}{c}{mean field} \\ \hline$R_c$ & 2.16 & 0.03 & 4.10 & 0.02 & 5.96 & 0.02 & 0.79788456 \\
$H_c$ & 1.435 & 0.004 & 1.265 & 0.007 & 1.175 & 0.004 & 0 \\
$B$ & 0.39 & 0.08 & 0.46 & 0.05 & 0.23 & 0.08 & 0 \\
\end{tabular}\vspace{0.25cm}
\caption{{\bf Non-universal scaling variables.}
Numerical values for the critical disorders and fields, and the
rotation parameter $B$ (equation~\ref{eq:scaling_2}),
in $3$, $4$, and $5$ dimensions extracted from scaling collapses. The
critical disorder is obtained from collapses of the spanning avalanches
and the second moments of the avalanche size distribution. The critical
field is obtained from the binned avalanche size distribution and the
magnetization curves. $H_c$ is affected by finite sizes, and systematic
errors could be larger than the ones listed here. The mean field
values are calculated analytically\protect\cite{prl1,Dahmen1}. The
rotation $B$ is obtained from the $dM/dH$ collapses. 
}
\label{table:RH}
\end{table}

\twocolumn

\section{Comparison with the analytical results}

Here we compare the simulation results with the renormalization group
analysis of the same system~\cite{Dahmen1,Dahmen2}. According to the
renormalization group the upper critical dimension (UCD), at and above
which the critical exponents are equal to the mean field values, is six.
Close to the UCD, it is possible to do a $6-\epsilon$ expansion,
and obtain estimates for the
critical exponents and the magnetization scaling function, which can
then be compared with our numerical results.

\begin{figure}
  \begin{center}
    \epsfxsize=8cm
    \epsffile{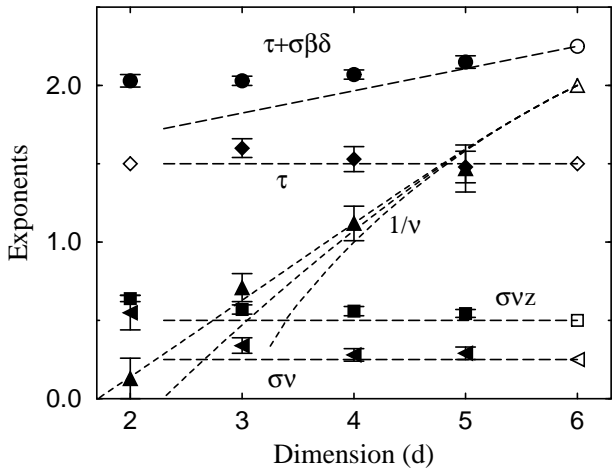}
  \end{center}
\caption{{\bf Comparison between the critical exponents from the
simulation and the $\epsilon$ expansion} Numerical values (filled
symbols) of the exponents $\tau + \sigma\beta\delta$, $\tau$, $1/\nu$,
$\sigma\nu z$, and $\sigma\nu$ (circles, diamond, triangles up, squares,
and triangle left) in $2$, $3$, $4$, and $5$ dimensions. The empty
symbols are values for these exponents in mean field (dimension 6).
Exponents in two dimensions are discussed
elsewhere~\protect\cite{prl3,OlgaLong,Kuntz}. Note that the value of $\tau$ in
$2$d conjectured value~\protect\cite{prl3}. We have simulated sizes up to
$30000^2$, $1000^3$, $80^4$, and $50^5$, where for $320^3$ for example,
more than $700$ different random field configurations were measured. The
long-dashed lines are the $\epsilon$ expansions to first order for the
exponents $\tau + \sigma\beta\delta$, $\tau$, $\sigma\nu z$, and
$\sigma\nu$. The short-dashed lines are Borel
sums\protect\cite{LeGuillou-Kleinert} for $1/\nu$, as discussed
in~\protect\cite{prl3}. The lowest is the variable-pole Borel sum from LeGuillou
{\it et al.}\protect\cite{LeGuillou-Kleinert}, the middle uses the
method of Vladimirov {\it et al.} to fifth order, and the upper uses the
method of LeGuillou {\it et al.}, but without the pole and with the
correct fifth order term. The error bars denote systematic errors in
finding the exponents from extrapolation of the values obtained from
collapses of curves at different disorders $R$. Statistical errors are
smaller.}
\label{fig:exp_compare}
\end{figure}

Figure~\ref{fig:exp_compare} shows the numerical and analytical results for
five of the critical exponents obtained in dimensions two to six (in six
dimensions, the values are the mean field ones). The other exponents can
be obtained from scaling relations\cite{Dahmen1}. The exponent
values in figure~\ref{fig:exp_compare} are obtained by extrapolating the
results of scaling collapses to either $R \rightarrow R_c$ or $1/L
\rightarrow 0$ (see~\cite{OlgaLong}). 
The long-dashed lines are the $\epsilon$ expansions to first
order for $\tau + \sigma\beta\delta$, $\tau$, $\sigma\nu z$, and
$\sigma\nu$. The three short-dashed lines\cite{Dahmen1} are Borel
sums\protect\cite{LeGuillou-Kleinert} for $1/\nu$. 
Notice that the numerical values converge
nicely to the mean field predictions, as the dimension approaches six,
and that the agreement between the numerical values and the $\epsilon$
expansion is quite impressive.

\begin{figure}
  \begin{center}
    \epsfxsize=8cm
    \epsffile{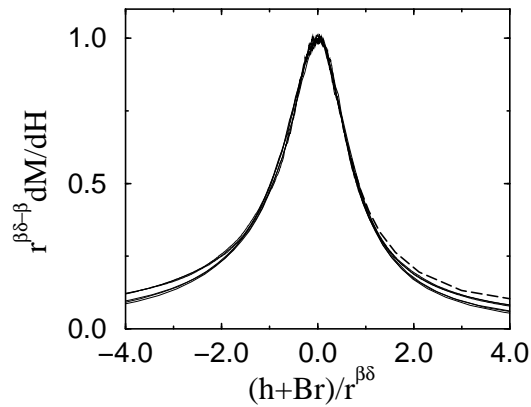}
  \end{center}
\caption{{\bf Comparison between simulated $dM/dH$ curves in $5$
dimensions, and the $dM/dH$ curve obtained from the $\epsilon$ expansion.}
The thick dashed line shows the prediction of the $\epsilon$ expansion
to third order in $\epsilon$ for the slope of the magnetization curve
$dM/dH$ in five dimensions. The theoretical curve is a parametric
form~\protect\cite{Zinn-Justin} taken from the analysis of the ordinary,
pure, thermal Ising model in three dimensions~\protect\cite{Dahmen2}. The 
six simulation curves (thin lines) are for a system of $30^5$ spins
at disorders $7.0, 7.3,$ and $7.5$ ($R_c=5.96$ in $5$ dimensions),
and for a system of $50^5$ spins at disorders $6.3, 6.4$, and $6.5$.
The latter curves are closer to the theoretical dashed line).
All the curves have been stretched/shrunk in the horizontal and vertical
direction and shifted horizontally to lie on each other.}
\label{fig:5d_dmdh}
\end{figure}

The $\epsilon$ expansion can be an even more powerful tool if it can
predict the scaling functions. This has been done for the magnetization
scaling function of the pure Ising model in $4-\epsilon$
dimensions~\cite{DombWallace,Zinn-Justin}. Since the $\epsilon$ expansion
for our model is the same as the one for the {\it equilibrium}
RFIM~\cite{Dahmen1}, and the latter has been mapped to {\it all} orders in
$\epsilon$ to the corresponding expansion of the regular Ising model in
two lower dimensions~\cite{Dahmen1,Aharony,Parisi}, we can use the
results obtained in~\cite{DombWallace,Zinn-Justin}. This is done in
figure~\ref{fig:5d_dmdh}, which shows the comparison between the
$dM/dH$ curves obtained in $5$ dimensions and the predicted
scaling function for $dM/dH$, to third order in $\epsilon$, 
where $\epsilon =1$ in $5$ dimensions (see~\cite{Zinn-Justin}).
As we see, the agreement is very good in the scaling region (close to
the peaks).

\section{Summary}

We have used the zero temperature random field Ising model, with a
Gaussian distribution of random fields, to model a random system that
exhibits hysteresis. We found that the model has a transition in the
shape of the hysteresis loop, and that the transition is critical. The
tunable parameters are the amount of disorder $R$ and the external
magnetic field $H$. The transition is marked by the appearance of an
infinite avalanche in the thermodynamic system. Near the critical point,
($R_C$, $H_C$), the scaling region is quite large: the system can
exhibit power law behavior for several decades, and still not be near
the critical transition. This is important to keep in mind whenever
experimental data are analyzed: decades of scaling need not imply
self--organized criticality.

We have extracted critical exponents for the magnetization, the
avalanche size distribution (integrated over the field and binned in the
field), the moments of the avalanche size distribution, the avalanche
correlation, the number of spanning avalanches, and the distribution of
times for different avalanche sizes. These values are listed in
Table~\ref{table:exponents}, and were obtained as an average of the
extrapolation results (to $R \rightarrow R_c$ or $L \rightarrow \infty$)
from several measurements~\cite{OlgaLong}. 
%For example,
%the correlation length exponent $\nu$ is the average value from three
%different collapses: the correlation function, the spanning avalanches,
%and the second moments of the avalanche size distribution, while the
%critical disorder $R_c$ is estimated from both the spanning avalanches
%collapses and the collapses of the moments of the avalanche size
%distribution. 
As shown earlier, the numerical results compare well with
the $\epsilon$ expansion~\cite{Dahmen1,Dahmen2}.  Comparisons to
experimental Barkhausen noise measurements~\cite{prl3} are very
encouraging.

We acknowledge the support of DOE Grant \#DE-FG02-88-ER45364 and NSF
Grant \#DMR-9805422. We thank Matt Kuntz for noticing and fixing the differences
between our analytical and numerical avalanche correlation functions.
We would like to thank Sivan Kartha, Bruce Roberts, M.
E. J. Newman, J. A. Krumhansl, J. Souletie, and M. O. Robbins for
helpful conversations. This work was conducted partly on the SP1 and SP2
at the Cornell Theory Center (funded in part by the National Science
Foundation, by New York State, and by IBM), and partly on IBM 560
workstations and an IBM J30 SMP system (both donated by IBM). We would
like to thank CNSF and IBM for their support. Further pedagogical
information using Mosaic is available at
http://www.lassp.cornell.edu/sethna/hysteresis/ and
http://SimScience.org/crackling/.


\begin{references}
\bibitem{note4} In $3$ and $4$ dimensions, below the critical field
$H_c$, we have not found any spanning avalanches. In $5$ dimensions,
spanning avalanches appeared below the critical field $H_c=1.175$. We believe
this is due to the small system size ($30^5$) used for the $5$
dimensional simulation. We however, do not expect substantial
corrections to the exponents derived from that data.
\end{references}
\end{document}